\documentclass[12pt,a4paper]{amsart}

\usepackage[all]{xy}
\usepackage{amssymb}
\makeatletter
\renewcommand{\@begintheorem}[2]{
\rm \trivlist \item [\hskip \labelsep {\bf #2\ \ #1.}]
                                }
\makeatother

\makeatletter

\DeclareFontFamily{U}{cyr}{}
\DeclareFontShape{U}{cyr}{m}{n}{
  <5> wncyr5 <6> wncyr6 <7> wncyr7 <8> wncyr8 <9> wncyr9 <10->
wncyr10}{}
\DeclareMathAlphabet{\mathcyr}{U}{cyr}{m}{n}

\input cyracc.def
\input epsf

\newcommand{\ZZ}{{\bf Z}}
\newcommand{\QQ}{{\bf Q}}

\newcommand{\CC}{{\bf C}}

\newcommand{\PP}{{\bf P}}

\newcommand{\cL}{{\mathcal L}}

\newcommand{\cO}{{\mathcal O}}

\newcommand{\bes}{\begin{equation*}}
\newcommand{\ees}{\end{equation*}}

\title{A new CY elliptic fibration and tadpole cancellation}

\author{Sergio L.~Cacciatori}
\address{Dipartimento di Scienze Fisiche e Matematiche, Universit\`a dell'Insubria,
Via Valleggio 11, I-22100 Como, Italia, and INFN, Sezione di Milano, Via Celoria 16, I-20133 Milano, Italia}
\email{sergio.cacciatori@uninsubria.it}
\author{Andrea Cattaneo}
\address{Dipartimento di Matematica, Universit\`a di Milano,
Via Saldini 50, I-20133 Milano, Italia}
\email{andrea.cattaneo1@unimi.it}
\author{Bert van Geemen}
\address{Dipartimento di Matematica, Universit\`a di Milano,
Via Saldini 50, I-20133 Milano, Italia}
\email{geemen@mat.unimi.it}

\begin{document}

\begin{abstract}
Tadpole cancellation in Sen limits in F-theory was recently studied by Aluffi and Esole. We extend their results, generalizing the elliptic fibrations they
used and obtaining a new case of universal tadpole cancellation, at least numerically. We could not find an actual Sen limit having the correct brane content,
and we argue that such a limit may not exist. We also give a uniform description of the fibration used by Aluffi and Esole as well as a new, simple, fibration
which has non-Kodaira type fibers.
\end{abstract}

\maketitle

\section*{Introduction}
Over the last years, F-theory has provided the most promising
connections between string theories and particle physics
phenomenology \cite{BHV,BHV1,HV,HKSV,HTV}. It allows a good
tunability of the parameters in order to generate realistic physical
scenarios, permits the stabilization of the moduli at high mass
scales, and maintains a classical Calabi-Yau geometry also in
presence of nontrivial fluxes. The Klebanov-Stressler mechanism can
be implemented in order to face the hierarchy problem. Cosmological
models can be accommodated, and the D-branes engineering includes a
natural unification. See \cite{D,H} for a review.

However, the theory suffers the drawback of being intrinsically non
perturbative, thus making the connection to weak coupling
perturbative physics quite hard to realize. Indeed, F-theory has
been proposed by Vafa \cite{V} as the strong coupling version of
type IIB strings, in a similar way as M-theory is the strong
coupling limit of type IIA strings. But whereas S-duality, which
interchanges low and strong couplings, maps IIA strings to M-theory
and vice versa, IIB string theory is self S-dual. The dynamics of
IIB strings is governed by the axion-dilaton field
$$
\tau=C_{(0)} +i\ e^{-\phi}.
$$
When this field acquires a constant vacuum expectation value (vev),
the string coupling constant is determined by the dilaton as
$g_s=e^{\phi}$. So a weak coupling corresponds to a large value of
the imaginary part of the axion-dilaton field: $Im\ \tau =g_s^{-1}$.
S-duality acts on this field as $SL(2,\mathbb{Z})$ in the usual
modular way
$$
\tau \longrightarrow \frac {a\tau +b}{c\tau+d}, \qquad\
\begin{pmatrix} a & b \\ c & d  \end{pmatrix} \in SL(2, \mathbb{Z}).
$$
The idea of Vafa has been to interpret $\tau$ as the modulus of an
elliptic curve in the same way as scalar fields can appear from the
moduli space of compactification varieties in higher dimensional
theories. In this case the extra dimensions should then correspond
to a family of elliptic curves parameterized by $\tau(x)$, $x\in
M^{1,9}$, the ten dimensional spacetime. One should be able to
recover type IIB strings from the compactification over the elliptic
curves of a twelve dimensional theory defined over an elliptic
fibration, which is F-theory \cite{V}. The actual construction is
not so simple because it is not possible to construct a twelve
dimensional supergravity with the right Lorentzian signature, so
that a chain of dualities must replace the naive twelve dimensional
construction. F-theory is thus realized by starting from M-theory
over an elliptic fibration. S-duality relates the M-theory to IIA
strings on the reduction over one of the cycles of the fibers.
T-duality next maps to IIB strings compactified over a circle and
finally recovers Lorentz invariance in the large radius limit. We
refer to \cite{D} and references therein for more details, and only
note that from this construction it follows that IIB string theory
compactified on a CY threefold $X$ is equivalent to F-theory over an
elliptically fibred CY fourfold $Y$. In general, the base $B$ of the
fibration is not necessarily a CY threefold, but $X$ can be obtained
as a double covering $\rho: X \rightarrow B$ of $B$.

The low energy physics is obtained by including D7- and D3-branes
filling the extended spacetime direction and then wrapping
four-dimensional cycles and points in the compact fourfold $Y$. The
presence of the branes implies the generation of fluxes that give
rise to tadpoles in general, as the total flux should vanish along
the compact directions. The tadpoles can be canceled by adding extra
fluxes that, however, in general break supersymmetry. In order to
save supersymmetry, it is convenient to allow only brane
configurations satisfying vanishing tadpole conditions. This is part
of the problem: as we said, taking the weak coupling limit
corresponds to considering a very large imaginary part of the
elliptic modulus $\tau$. However, in F-theory, $\tau$ is not
constant so that condition can be true for example at the position
of a single brane. We can choose which brane can be tuned at the
weak coupling limit by means of an S-duality transformation, but
this cannot be done for all branes simultaneously. Thus, the weak
coupling limit can generically be performed only locally and not
globally. This makes computability in F-theory quite a problem from
the perturbative point of view.

In order to recover perturbative type IIB string theory on $X$ one
has to take a limit, in F-theory, such that the axion-dilaton field
becomes constant along the whole base $B$, with divergent imaginary
part. Thus the elliptic fibration acquires an O7-orientifold
singularity in $B$ and the $D7$ branes collapse on the orientifold
in $X$. This is the Sen limit \cite{S1,S}. It is still the only
convincing mechanism to get the limit. Recently Sen limits were
studied in \cite{AE} and \cite{BHT}.

This paper is in a sense a sequel (or maybe a footnote) to the paper
\cite{AE} by Aluffi and Esole. In that paper important new insights
into Sen limits and tadpole cancellation were obtained. Moreover,
new examples of tadpole cancellation were found.

These new examples of tadpole cancellation show that it is important
to have detailed descriptions of elliptic fibrations and their
degenerations. In this paper we introduce a new family which from
many points of view is similar to the $E_i$, $i=6,7,8$, families
used in in \cite{AE}. As we show in section \ref{ne7}, these three
families are all realized as subvarieties of  projective plane
bundles $\PP E$, where $E$ is a direct sum of three line bundles on
the base. Our description of the $E_7$ fibration in particular is
simpler than the original one.

The new family is also of this type. This led us to compute the
Chern classes for fibrations of this type in a uniform manner, using
and generalizing the methods from \cite{AE}. These allow us to study
universal (i.e.\ independent of the base) tadpole cancellation in
Sen limits in a uniform manner. The results are, in a sense,
disappointing: there is only one new case where one could have
tadpole cancellation. The fibration involved happens to be the new
fibration we introduced. However, it seems to be impossible to get
the correct brane configuration in the Sen limit which is required
for tadpole cancellation. It would be interesting to understand if
this means that this configuration is thus intrinsically strongly
coupled or if there exists a generalization of Sen's
mechanism which provides a geometric realization of the weak coupling limit for this new example.\\
It is worth to mention that in \cite{GW} the authors get a formula
for the induced tadpole of a certain class of spectral cover models
which is checked against the explicitly computed tadpoles of the
fourfolds in their examples. In \cite{GW2} the relevance of the
global properties of the fibration for the induced $D3$-brane
tadpole condition has been put in evidence.

The new family has another feature: even when the total space of the
fibration is smooth, it can have singular fibers which are not of
Kodaira type whenever the base has dimension at least two. This is
not a contradiction to anything, as Kodaira's classification applies
only to elliptic fibrations over a curve. Till recently (in
particular, before the appearance of  \cite{EY}), such fibers have
not received much attention and they are still problematic in
F-theory. For example, in \cite{MSS1,MSS2} and \cite{DMSS} the
authors look for suitable geometric constructions in order to avoid
non-Kodaira fibers. Indeed, in such fibers (which in general can
appear over points or along curves in the three dimensional base)
the spectral cover becomes singular and the usual techniques for
reading the physical spectrum are not available anymore. Note also
that a parallel construction of well-defined Calabi-Yau fourfolds
for F-theory compacitifications has been carried out in \cite{GW},
\cite{GW1}. In these papers, explicit examples of singular elliptic
fourfolds and their full resolution are obtained as complete
intersections in toric spaces. In fact, there the authors point out
the importance of the resolution as the way to check consistency of
the fourfold, which a priori is by no means to be taken for granted
just by applying the Kodaira reasoning to a higher dimensional base
space. Therefore, we hope that a better understanding of non-Kodaira
fibers will lead to new insights in physics. The fact that our new
family is so simple (it is defined by just one  equation in a
$\PP^2$-bundle) and that is has an even simpler local equation near
a non-Kodaira fiber, should make it very suitable for further study.

In a sense, our family is not new at all. If the base is a toric
variety, then the projective plane bundle is also a toric variety.
Calabi-Yau hypersurfaces of toric varieties have been studied
extensively in the last decades. As far as we know, no special
attention has been paid to this example till now however.

In the Appendix we review Sen's limit. The limit fibration is shown
to have a natural orbifold interpretation. It has a double cover
which is the product of the Calabi-Yau double cover of the base and
a fixed singular elliptic curve. In Sen's example of a weak coupling
limit, the local monodromy around two $I_1$ fibers is the inverse of
the one of an $I_4^*$ fibre. The monodromy is essential for
identifying the branes in the weak coupling limit. We give a toy
model which has this monodromy. It is a fibration over the
projective line with three singular fibers.

\

\section{A new model for the $E_7$ family}\label{ne7}

Aluffi and Esole showed in \cite{AE} that certain elliptic
fibrations, the $E_6$, $E_7$ and $E_8$ families, are quite useful in
the study of tadpole cancellation. The $E_6$ fibration is defined by
a cubic equation in $\PP(\cO\oplus \cL\oplus\cL^{})$, cf. \cite{AE},
eqn.\ (1) and section \ref{e6tp} below. The $E_8$ fibration can also
be given by the Weierstrass model
$$
y^2z=x^3+fxz^2+gz^3\quad\mbox{in}\quad \PP(\cO\oplus\cL^{\otimes
2}\oplus\cL^{\otimes 3}), \quad f\,\in H^0(B,\cL^{\otimes 4}),\quad
g\,\in H^0(B,\cL^{\otimes 6})
$$
(although in equation (3) of \cite{AE} a different ambient bundle is
chosen). We show that the $E_7$ families are hypersurfaces in a
$\PP^2$ bundle  $\PP(\cO_B\oplus \cL\oplus \cL^{\otimes 2})$, just
like the $E_6$ and $E_8$ family. This uniform description of these
three fibrations is quite helpful in (re)checking the tadpole
cancellation, see section \ref{tpcc1}.

\subsection{The $E_7$ family}
Let $B$ be a complex manifold, and let $\cL$ be a line bundle on
$B$. An $E_7$ fibration over $B$ is defined by a quartic equation
$$
y^2\,=\,x^4\,+\,ex^2z^2\,+\,fxz^3\,+\,gz^4
$$
with $(z:x:y)$ coordinates on $Z:=\PP_{1,1,2}(\cO_B\oplus \cL\oplus
\cL^{\otimes 2})$. The coefficients are global sections of the line
bundles:
$$
e\,\in H^0(B,\cL^{\otimes 2}),\qquad f\,\in H^0(B,\cL^{\otimes
3}),\qquad g\,\in H^0(B,\cL^{\otimes 4}),
$$
and the coordinates $x,y,z$ are sections of the line bundles $\cL
\otimes \cO_Z(1)$, $\cL^{\otimes 2} \otimes \cO_Z(2)$ and $\cO_Z(1)$
respectively.

A quartic equation actually has more terms, as one can add
$x^2y,xyz,x^3z,z^2y$ with suitable coefficients. However, a general
quartic equation can be transformed in the one above. If one chooses
$\cL=\omega_B^{-1}$, the equation defines a Calabi-Yau variety.

\subsection{A $\PP^3$ bundle}
The weighted projective plane $\PP_{1,1,2}$ is isomorphic to a
quadric cone in $\PP^3$. The $E_7$ fibration can also be obtained as
the intersection of a fixed (singular) quadric fibration with
another  quadric fibration inside a $\PP^3$-bundle over $B$. The
isomorphism is induced by
$$
\PP_{1,1,2}(\cO_B\oplus \cL\oplus \cL^{\otimes
2})\,\longrightarrow\, \PP(\cO_B\oplus \cL\oplus \cL^{\otimes 2}
\oplus \cL^{\otimes 2}),
$$
$$
(z:x:y)\,\longmapsto\,(X_0:\ldots:X_3)\,:=\,(z^2:xz:x^2:y).
$$
The image of the map is defined by two quadrics
$$
X_0X_2\,-\,X_1^2\,=0,\qquad
X_3^2\,=\,X_2^2\,+\,eX_1^2\,+\,fX_0X_1\,+\,gX_0^2,
$$
the first defines the image of the weighted projective plane
bundles.

\subsection{The new model}
We project this fibration from the (constant) section $(0:0:1:1)$ to
a $\PP^2$ bundle. Such a projection is
$$
\PP(\cO_B\oplus \cL\oplus \cL^{\otimes 2}\oplus \cL^{\otimes 2}) \,-
- - \rightarrow\,\PP(\cO_B\oplus \cL\oplus \cL^{\otimes 2}),
$$
where, recycling the coordinates $x,y,z$:
$$
(X_0:X_1:X_2:X_3)\,\longmapsto\,(z:x:y)\,:=\,(X_0:X_1:X_2-X_3).
$$

The image of the intersection of the two quadric bundles is the
family of cubic curves
$$
gz^3\,+\,fz^2x\,+\,ezx^2\,+\,2x^2y\,-\,zy^2\,=\,0\qquad(\subset
\PP(\cO_B\oplus \cL\oplus \cL^{\otimes 2})).
$$
Here $z,x,y$ are global sections of $\cO_W(1)$, $\cL\otimes\cO_W(1)$
and $\cL^{\otimes 2}\otimes\cO_W(1)$ respectively, where
$W:=\PP(\cO_B\oplus \cL\oplus \cL^{\otimes 2})$. This is our new
description of the $E_7$ family.

The inverse of the projection on this curve is the map
$$
(z:x:y)\,\longmapsto\,(z^2:xz:x^2:x^2-yz).
$$
We refer to \cite{Cas}, $\S$8 for these computations.

\section{A new family}\label{nkf}

As the dimension of the base is typically three for applications to
F-theory, Kodaira's standard classification of singular fibers of an
elliptic fibration over a curve can no longer be applied. New fiber
types can arise over codimension two subvarieties of the base. In
this section we provide an example with non-Kodaira type fibers. The
associated physics is not yet clear to us, but see the recent
preprint \cite{EY}. In section \ref{tpcc1} we will see that this
family is of particular interest for a strong form of tadpole
cancellation.

\subsection{Equations}\label{E5eq}\label{wY}
Our new example consists of the elliptic fibrations
$\phi:Y\rightarrow B$ defined by a cubic equation $F=0$ in the
$\PP^2$-bundle $Z$ defined in (\ref{eqY}), with coordinates
$(x:y:z)$, where
$$
F\,=\,a_0x^2y+a_1x^2z+b_0xy^2+b_1xyz+b_2xz^2+
c_0y^3+c_1y^2z+c_2yz^2+c_3 z^3
$$
with $a_i\in H^0(B,\cO)$, $b_i\in H^0(B,\cL)$ and $c_i\in
H^0(B,\cL^{\otimes 2})$,
\begin{eqnarray}\label{eqY}
Y:\quad (F=0)\qquad\subset\,Z:=\,\PP E,\qquad E:=\cL\oplus
\cO_B\oplus \cO_B,
\end{eqnarray}
$x$ is a section of $\cL \otimes \cO_Z(1)$ and $y$, $z$ are sections
of $\cO_Z(1)$. Let $H$ be the class of $\cO_Z(1)$, let $L$ be the
class of $\cL$ and let $p:Z\rightarrow B$ be the projection. Then
$Y$ has class $3H+2p^*L$.

In case $a_0=a_1=0$, the variety $Y$ is singular in
$(x:y:z)=(1:0:0)$ (for all $b\in B$). Assuming $Y$ to be smooth, we
may thus assume that $a_0\neq 0$ (else we permute $y$ and $z$). The
projective transformation
\begin{eqnarray*}
  && x=x'-\frac {b_0}{2a_0^2} y'-\frac {a_0 b_1-2 a_1 b_0}{2 a_0^2} z', \\
  && y=\frac 1{a_0} y' - \frac {a_1}{a_0} z', \\
  && z=z',
\end{eqnarray*}
gives the new equation
\begin{equation*}
F\,=\,x^{\prime2}y'+b'_2x'z^{\prime2}+
c'_0y^{\prime3}+c'_1y^{\prime2}z'+c'_2y'z^{\prime2}+c'_3z^{\prime3}.
\end{equation*}
Thus, after dropping all primes, we may assume that the equation
defining a smooth $Y$ is of the simpler form
\begin{equation}\label{stdF}
F\,:=\,x^2y+b_2xz^2+ c_0y^3+c_1y^2z+c_2yz^2+c_3z^3.
\end{equation}

The fibration $Y$ defined by $F$ has a (constant) section
$$
S\,:\;B\,\longrightarrow\, Y,\qquad b\,\longmapsto \,((1:0:0),b).
$$
This allows one to put the equation in Weierstrass form, for example
using the algorithm given in \cite{Cas}. A Weierstrass form is
$y^2=x^3+fx+g$ with {\renewcommand{\arraystretch}{1.3}
\begin{eqnarray*}
f&=&36b_2^2c_0 + 36c_1c_3 - 12c_2^2,\\
g&=&-144b_2^2c_0c_2 + 54b_2^2c_1^2 - 216c_0c_3^2 + 72c_1c_2c_3 -
16c_2^3.
\end{eqnarray*}}
In general, the discriminant $4f^3+27g^2$ is rather complicated, but
in case $b_2=0$ it reduces to the discriminant of the binary cubic
form $c_0y^3+c_1y^2z+c_2yz^2+c_3z^3$.

\subsection{Smooth fibrations}\label{smnkf}
In case one has a smooth fibration, defined by $F=0$, any small
deformation of the coefficients $b_i,c_j$ will also be a smooth
fibration. We will show that if $c_0=0$ and $c_1=0$ define two
smooth subvarieties of $B$ which intersect transversely, then
\begin{eqnarray} \label{eqG}
G&=&x^2y+c_0y^3+c_1(y^2z+z^3)
\end{eqnarray}
defines a smooth fibration.

In fact, assume that $q:=(x:y:z)\in p^{-1}(b)$ is a singular point.
The $x$-derivative shows that $xy=0$ hence $x=0$ and
$c_0y^3+c_1(y^2z+z^3)=0$ or we have $y=0$ and $c_1z^3=0$. In the
last case, if $y=z=0$ we get $q=(1:0:0)$, which is always a smooth
point however. Thus if $y=0$ we must have $c_1=0$. The
$y$-derivative is $x^2+3c_0y^2+2c_1yz$ so if $y=0$ also $x=0$ and
$q=(0:0:1)\in p^{-1}(b)$ with $c_1(b)=0$. As $c_1=0$ is assumed to
be smooth, there is a (local) derivation $\partial$ on $B$ such that
$(\partial c_1)(b)\neq 0$. Since $(\partial G)(q)=(\partial c_1)(b)$
it follows that a singular point has $y\neq 0$ and $x=0$. If $z=0$
we would have $c_0(b)=0$ and $(\partial c_0)(b)y^3=0$ which
contradicts smoothness of $c_0=0$. Thus $y,z\neq 0$ and we may
assume $q=(0:y:1)$.

The $z$-derivative is $c_1(y^2+3z^2)$, so if $c_1(b)\neq 0$ then
$y=\pm\sqrt{-3}$. Now the $x$ and $y$ derivatives in $q$ show that
$c_0(b)y^3+c_1(b)(y^2+1)=0$ and $3c_0(b)y+2c_1(b)z=0$, which is a
contradiction. So a singular point of $Y$ is a point of a fiber over
$b\in(c_0=c_1=0)$. These varieties intersect transversely, hence
there are derivations $\partial_i$ on the base, $i=1,2$, where the
matrix $(\partial_i c_j)(b)$ has rank two and thus, for at least one
of these derivations, $(\partial G)((0:y:1),b)\neq 0$.

\subsection{Calabi-Yau fibrations}\label{cyfibs}
The family $Y$ will be a Calabi-Yau variety if its class $3H+2p^*L$
in $Z$ is the anticanonical class $-K_Z$ of $Z$. As
$-K_Z=3H+p^*L+p^*c_1(T_B)$ (cf.\ \cite{F}, B.5.8), this occurs if we
choose $L=c_1(B)$. Thus if $\cL\cong\omega_B^{-1}$, the inverse of
the canonical line bundle on $B$, we get Calabi-Yau fibrations.

For example, if $B=\PP^n$ we take $\cL=\cO(-n-1)$. Sections
$c_0,c_1\in H^0(\PP^n,\cO(2n+2))$ which give smooth, transversally
intersecting subvarieties are for example $c_0=\sum x_i^{2n+2}$ and
$c_1=\sum 2^ix_i^{2n+2}$ and these give smooth CY fibrations over
$\PP^n$ with equation $G$ as in formula \ref{eqG}.

\subsection{Non-Kodaira fibers}
We consider again the elliptic fibration $\phi:Y\rightarrow B$ given
by the equation (\ref{eqG}) from section \ref{smnkf} and we assume
that $Y$ is smooth. The fiber of $\phi$ over a point $b\in B$ in the
codimension two subvariety  $c_0(b)=c_1(b)=0$ is defined by
$x^2y=0$. It is a singular fiber with two irreducible components,
which does not occur in Kodaira's list. In fact, Kodaira listed the
possible singular fibers in case the base has dimension one. For
higher dimensions his methods still apply to the generic points of
codimension one subvarieties, but they do not extend to codimension
two subvarieties of the base. So, a priori, there is no reason not
to expect non-Kodaira fibers over codimension two subvarieties. A
fibration with non-Kodaira fibers appeared in \cite{GGO}, but their
nature and physical significance is not clear. We wonder if there is
interesting physics associated to non-Kodaira fibers. We will not
consider this aspect here, leaving it for future work. See also the
recent preprint \cite{EY} for non-Kodaira fibers.

Assuming that $c_0=0$ and $c_1=0$ are smooth and intersect
transversally,  we can use $c_0,c_1$ as part of a coordinate system
in a point of intersection. So locally on $B$ we are basically
dealing with the two parameter family
$$
x^2y+sy^3+t(y^2z+z^3)\,=\,0\qquad (\subset
\PP^2_{(x:y:z)}\times\CC^2_{(s,t)})
$$
near $(s,t)=(0,0)$. The threefold defined by this equation is
smooth, the projection to $\CC^2$ is an elliptic fibration with
(constant) section $(1:0:0)$ and the fiber over $s=t=0$ is not of
Kodaira type. The discriminant and j-invariant are
$$
\Delta\,=\,t^4(4t^2+27s^2),\qquad j\,=\,6912\frac{t^2}{4t^2+27s^2}.
$$
For $(s,t)\neq (0,0)$, the singular fibers are of type $IV$ (three
concurrent lines) if $t=0$, and they are nodal cubics if
$4t^2+27s^2=0$ , this subvariety is the union of two lines defined
by $2t\pm \sqrt{-27}s=0$. Thus the non-Kodaira fiber arises from a
collision of three singular Kodaira fibers. In \cite{M} it is shown
that collisions can be avoided at the cost of repeatedly blowing up
the base (which will in general imply that a CY fibration is
modified in a non-CY fibration). The fibers of the modified
fibration will be of Kodaira type.

One should notice that when we restrict the fibration to a smooth
curve passing through $(0,0)$, the fibration restricts to give a
singular surface (in fact, else we \emph{would} have a Kodaira fiber
over $(0,0)$). To see this, let us consider for simplicity a line
through $(0,0)$ parametrized by $(s,t)=(au,bu)$. The family
restricts to the surface defined by $x^2y+u(ay^3+b(y^2z+z^3))=0$
over this line. It is singular in the points $(u,x,y)$ with
$u=x=ay^3+b(y^2z+z^3)=0$. For general $(a,b)$ one gets three
singular points and the surface is minimally desingularized by
blowing them up. The resulting Kodaira fiber has type $I_0^*$. Note
that the position of the singular points on the surface depends on
$a,b$.

To deal with all possible $(a,b)$, one might guess that one could
blow up the local family in the line $u = x = 0$, but doing so will
introduce an exceptional divisor of dimension 2 which will be in the
fiber over the point $(0,0)$. So the fiber dimension jumps and we do
not have an elliptic fibration anymore.

\subsection{Remark} The second author has obtained various other results on CY elliptic fibrations $Y$ in $\PP(\cO\oplus \cL_1\oplus\cL_2)$ over a base $B$.
In particular, for $B=\PP^2$, all possible line bundles $\cL_i$, the
cohomology of $Y$  and the cubic intersection form on $Pic(Y)$ have
been determined. These results will appear elsewhere.

\section{Chern classes}

\subsection{Motivation}\label{aer}
Let $\phi:Y\rightarrow B$ be an elliptic fibration. Aluffi and Esole
(\cite{AE}, Thm 4.3, see also \cite{AE2}, $\S$4.2 for the
$E_8$-case) found that the push-forward of the total Chern class
$c(T_Y)$ of $Y$ along the fibration  $\phi:Y\rightarrow B$ is equal
to the total Chern class of a codimension one subvariety $C_d$ in
$B$, up to an integer $m$, for certain fibrations:
 {\renewcommand{\arraystretch}{1.3}
$$
\phi_*c(T_Y)\,=\,m\cdot c(T_{C_d}), \qquad\qquad
\begin{array}{|c|c|c|c|}\hline
&E_8&E_7&E_6\\ \hline m&2&3&4\\ \hline d&6&4&3 \\ \hline
(d_1,d_2)&(2,3)&(1,2)&(1,1)\\ \hline
\end{array}
$$
}
here $C_d\subset B$ is a smooth hypersurface in $B$, with class
$dL$. Notice that in these three cases $d=d_1+d_2+1$ and that one
assumes that $c_1(\cL)=c_1(B)$. We will generalize this formula.
Although the result is not as concise, it will be very useful for
determining the Chern classes of $Y$ and for applications to tadpole
cancellation. The recent paper \cite{FJ} by J.\ Fullwood,  of which
we were unfortunately not aware when we wrote this paper, provides a
general result for dealing with $\phi_*$.

\subsection{Computing the Chern classes}\label{cec}
We recall the method used in \cite{AE} to obtain the results we just
recalled, and then we provide the generalization.

Let $Y$ be a (smooth) Calabi-Yau hypersurface in a bundle of
projective planes $Z:=\PP E$ where $E$ is a rank three vector bundle
over a base manifold $B$. We will assume that $E=\oplus \cL^{\otimes
d_i}$ is a direct sum of three line bundles on $B$. Replacing $E$ by
$E\otimes \cL^{-d_3}$ does not change $\PP E$ and we are in the case
$E\cong\cO\oplus\cL^{\otimes{d_1}}\oplus\cL^{\otimes d_2}$. We write
$d_iL:=c_1(\cL^{\otimes d_i})$ for the first Chern classes.
\[\xymatrix{
Y \ar[r]^{i} \ar[dr]_\phi  &Z\,=\PP E \ar[d]_p&\qquad
&E\,\cong\,\cO\,\oplus\,\cL^{\otimes{d_1}}\,\oplus\,\cL^{\otimes d_2},\\
 &  B&&L:=c_1(\cL),\quad H:=c_1(\cO_{\PP(E)}(1)).
} \] We will assume that $[Y]=-K_Z$ in $H^*(Z,\QQ)$, where $K_Z$ is
the canonical class of $Z$, this implies that $Y$ is Calabi-Yau.
Moreover, we assume that $c_1(B)$ is a multiple of $L$, we simply
write  $c_1(B)=c_1L$ for some $c_1\in\ZZ$. Then
$$
-K_Z=3H+(c_1+\sum d_i)p^*L,\qquad\mbox{with}\quad  c_1(B)=c_1L.
$$
The relative tangent bundle sequence on $Z$ gives:
$$
c(T_Z)\,=\,c(T_{Z/B})p^*c(T_B)\,=\, c(p^*E\otimes
\cO_{\PP(E)}(1))p^*c(T_B)\,=\,
\Bigl(\prod_{i=1}^3(1+d_ip^*L+H)\Bigr)p^*c(T_B).
$$
Slightly generalizing the computations in \cite{AE}, proof of
Theorem 4.3, omitting $i_*$ we get:
$$
c(T_Y)\,=\,\left(\frac{c(T_Z)}{1-K_Z}\right)_{|Y}\,=\,\frac{\prod_{i=1}^3(1+d_ip^*L+H)}
{1+3H+(c_1+\sum d_i)p^*L} \Bigr(3H+(c_1+\sum
d_i)p^*L\Bigl)p^*c(T_B).
$$
Note that we put $d_3=0$. The general result is that
\begin{eqnarray}\label{PQ}
\phi_*(T_Y)\,=\,\frac{P}{Q}c(T_B),
\end{eqnarray}
with polynomials {\renewcommand{\arraystretch}{1.3}
\begin{eqnarray*}
P&:=&3L(4c_1+P_2L+P_3L^2),\phantom{\Bigl(\Bigr)}\\
&&P_2\,:=\,2(-d_1^2+d_1d_2-d_2^2+c_1^2),\\
&&P_3\,:=\,-2d_1^3+3d_1^2d_2+3d_1d_2^2-2d_2^3-3(d_1^2-d_1d_2+d_2^2)c_1+c_1^3,\phantom{\Bigl(\Bigr)}\\
Q&:=&(1+(c_1+d_1-2d_2)L)(1+(c_1+d_1+d_2)L)(1+(c_1-2d_1+d_2)L).
\end{eqnarray*}
}

\

To obtain a formula more similar to the one from \cite{AE}, notice
that if  $c_1=1$, so $\cL\cong\omega_B^{-1}$, and
$d_1d_2(d_1-d_2)\neq 0$ (so $Q$ is a product of three distinct
factors), then this formula implies:
$$
\phi_*c(T_Y)\,=\, m_pc(T_{C_p})\,+\,m_qc(T_{C_q})\,+\,m_rc(T_{C_r})
\qquad \mbox{with}\quad c(T_{C_d})\,=\,\frac{dL}{1+dL}c(T_B)
$$
with integers:
$$
p\,:=\,-2d_1+d_2+1,\qquad q\,:=\,d_1+d_2+1,\qquad r\,:=\,d_1-2d_2+1,
$$
and rational numbers:
$$
m_p\,:=\,\frac{(d_1-1)(d_1-d_2-1)}{d_1(d_1-d_2)}, \quad
m_q:=\frac{(d_1+1)(d_2+1)}{d_1d_2}, \quad
m_r:=\,-\frac{(d_2-1)(d_2-d_1-1)}{d_2(d_1-d_2)}.
$$

The formulas from \cite{AE}, cf.\ section \ref{aer} above, can now
be derived by simply substituting $c_1=1$, $d_1=2$, $d_2=3$;
$c_1=1$, $d_1=1$, $d_2=2$ and $c_1$, $d_1=d_2=1$ in $P$, $Q$ for the
$E_8$, $E_7$ and $E_6$ case respectively.

The new family we introduced in section \ref{nkf} has $d_1=0,
d_2=1$, taking also $c_1=1$ we get
$$
\phi_*c(T_Y)\,=\, \frac{3L(4-4L^2)}{(1-L)(1+2L)^2}c(T_B)\,=\,
\frac{12L(1+L)}{(1+2L)^2}c(T_B).
$$

\subsection{Verifying the formula}
Recall the defining relation of the Chern classes of $E$, where
$H:=c_1(\cO_{\PP E}(1))$:
$$
H^3\,=\,-(p^*c_1(E)H^2\,+p^*c_2(E)H+p^*c_3(E)).
$$
Note that we don't have alternating signs since, following the
convention in physics, $\PP E$ denotes the projective bundle of one
dimensional subspaces in $E$ rather then codimension one subspaces.

For dimension reasons, $p_*1_Z=0$ and $p_*H=0$, moreover
$p_*H^2=1_B$ and more generally one can define the Segre classes
$s_i(E)$ of $E$ by
$$
s_i(E)\,=\,p_*H^{i+2}\qquad(\in H^{2i}(B,\QQ)).
$$

Using the projection formula, $p_*(ap^*b)=(p_*a)b$ for classes $a,b$
on $Z$ and $B$ respectively, we can compute $s_i(E)=p_*H^{i+2}$ by
induction from the defining relation:
 {\renewcommand{\arraystretch}{1.3}
$$
\begin{array}{rcl}
s_{i+1}(E)&=&-p_*(p^*c_1(E)H^{i+2}-p^*c_2(E)H^{i+1}+p^*c_3(E)H^i)\\
&=& -(c_1(E)s_{i}(E)+c_2(E)s_{i-1}(E)+c_3(E)s_{i-2}(E)).
\end{array}
$$
} 
This implies the following identity in the cohomology ring $\oplus_i
H^{2i}(B,\QQ)$:
$$
\left(\sum_{i=0}^\infty
s_i(E)\right)\bigl(1+c_1(E)+c_2(E)+c_3(E)\bigr)  \,=\,1.
$$

In our case, $E=\oplus \cL^{d_i}$ and $L:=c_1(\cL)$, so
$1+c_1(E)+c_2(E)+c_3(E)=\prod(1+d_iL)$ and for $j \geq 2$
$$
p_*H^{j}\,=\,n_{j-2}L^{j-2}\quad\mbox{where}\quad
\frac{1}{\prod(1+d_iL)}\,=\,\sum_{j=0}^\infty n_jL^j.
$$
The formula (\ref{PQ}) can now be derived with lengthy computations,
or, better, by using the recent results of J.\ Fullwood in
\cite{FJ}.

\subsection{Euler characteristics of elliptic fibrations}\label{chel}
In section \ref{cec} we expressed $\phi_*c(Y)$, the push-forward of
the total Chern class of $Y$ to $X$, as a rational function
multiplied by $c(T_B)$. Let $d=\dim Y$, then the Euler
characteristic $\chi(Y)$ of $Y$ is  equal to $\int_Yc_d(Y)$
(Gauss-Bonnet) which is the same as $\int_B \phi_*c_d(Y)$.

Thus $\chi(Y)$ is the coefficient of the term of degree $\dim B$ in
the product of $P/Q=1+a_1L+a_2L^2+\ldots$ with
$c(T_B):=1+c_1(B)+c_2(B)+\ldots$ (where $L$ has degree $1$ and
$c_i(B)$ has degree $i$). We found:
$$
\chi(Y)\,=\,-6((d_1^2 - d_1d_2 + d_2^2)L^2 +3c_1^2),
$$
in case $\dim B=2$, whereas for $\dim B=3$ we found (with
$c_i=c_i(B)$):
$$
\chi(Y)\,=\,27c_1^3 + 12c_1c_2+39(d_1^2 - d_1d_2 + d_2^2)c_1L^2+
3(-2d_1^3+3d_1^2d_2+3d_1d_2^2-2d_2^3)L^3.
$$
To recover the formula for the $E_7$ case from \cite{AE},
Proposition 4.2 we substitute $L=c_1$ and $(d_1,d_2)=(1,2)$, which
gives $\chi(Y)=12c_1(B)c_2(B)+144c_1(B)^3$. For the new family, with
$L=c_1$, we put $(d_1,d_2)=(0,1)$ and we get $\chi(Y)=-24c_1(B)^2$
and $\chi(Y)=12c_1(B)c_2(B)+60c_1(B)^3$ in case $\dim Y=3,4$
respectively.

\subsection{The push-forward from $X$ to $B$}\label{pf}
In order to check the tadpole relations, we need to understand the
push forward along a double cover $\rho:X\rightarrow B$ of Chern
classes of subvarieties of $X$. Here $X\subset \cL$ is a subvariety
of the total space of the line bundle $\cL$ defined by an equation
$\xi^2=h$, where $h\in H^0(B,\cL^{\otimes 2})$. We denote by
$\tilde{\rho}:\cL\rightarrow B$ the bundle projection, which
restricts to $\rho$ on $X\subset \cL$.
\[
\xymatrix{X \ar[r]^{i} \ar[dr]_\rho  &\cL \ar[d]_{\tilde{\rho}}\\ &
B}
\]

First of all we consider the Chern classes of $X$. From the normal
bundle sequence associated to $X\subset \cL$ we get
$c(T_X)=(c(T_\cL)/c(N_{X/\cL}))_{|X}$. As
$N_{X/\cL}\cong\cO_\cL(X)\cong \tilde{\rho}^*\cL^{\otimes 2}$, we
get $c(T_X)=(c(T_\cL)/c(\tilde{\rho}^*\cL^{\otimes 2}))_{|X}$. The
inclusion of the zero section of $\cL$, which we identify with $B$,
gives the relation $c(T_\cL)=\tilde{\rho}^*c(T_B)c(N_{B/\cL})$. As
$N_{B/\cL}=\cL$ one has
$c(T_\cL)=\tilde{\rho}^*c(T_B)(1+\tilde{\rho}^*L)$ thus:
$$
c(T_X)\,=\,\left(\frac{c(T_\cL)}{c(N_{X/\cL})}\right)_{|X}\,=\,
\left(
\frac{\tilde{\rho}^*c(T_B)(1+\tilde{\rho}^*L)}{1+2\tilde{\rho}^*L}\right)_
{|X}\,=\, \frac{1+\rho^*L}{1+2\rho^*L}\rho^*c(T_B).
$$
As $\rho$ is a 2:1 map, the projection formula implies that
$\rho_*\rho^*Z=2Z$ for any cohomology class $Z$ on $B$. Thus if
$D\subset X$ is a smooth hypersurface in $X$ of class $a\rho^*L$
then (cf.\ \cite{AE}, 4.3.3)
$$
\rho_*c(T_D)\,=\,\rho_*\left(\frac{a\rho^*L}{1+a\rho^*L}c(T_X)\right)\,=\,
\frac{2aL(1+L)}{(1+aL)(1+2L)}c(T_B).
$$

\

\

\section{Tadpole cancellation}

In \cite{AE}, Aluffi and Esole established general results on
tadpole cancellation. We recall one of their examples to illustrate
the general approach. See also the Appendix for more background on
Sen limits. In section \ref{tpcc1} we use our results on Chern
classes for elliptic fibrations in $\PP^2$-bundles to find all
numerical examples of generalized strong tadpole cancellation for
such fibrations in the case that $c_1(B)=c_1L$ for a rational number
$c_1$.

There are only three cases, two of which were already done in
\cite{AE}. The other case occurs for the new family we introduced in
section \ref{nkf}. However, we could not find a Sen limit of such a
family which has the correct limiting geometry and we argue that
such a family may not exist.

\subsection{The $E_6$-example from \cite{AE}}\label{e6tp}
The $E_6$-fibration (cf.\ \cite{AE}, $\S$2) over a complex manifold
$B$ is defined by an equation
$$
y^3\,+\,x^3\,-\,dxyz\,-\,exz^2\,-fyz^2\,-gz^3=0
$$
where $(z:x:y)$ are coordinates on the fibers of the projective
plane bundle $\PP (\cO_B\oplus\cL\oplus\cL)$,
$$
d\,\in\,H^0(B,\cL),\qquad e,f\,\in\,H^0(B,\cL^{\otimes 2}),\qquad
g\,\in\,H^0(B,\cL^{\otimes 3}).
$$
This family has three (constant) sections: $(z:x:y)=(0:1:\rho)$ with
$\rho^3=-1$.

In \cite{AE}, $\S$3.4 the following Sen limit is considered:
$$
\left\{\begin{array}{rcl}
d&=&6k\\
e&=&9k^2+3h\\
f&=&9k^2+3h+C\phi\\
g&=&2k(5k^2+3h)+C(\gamma+k\phi)
\end{array}\right.\qquad
\left\{\begin{array}{rcl}
k&\in& H^0(B,\cL)\\
h,\phi&\in& H^0(B,\cL^{\otimes 2})\\
\gamma&\in& H^0(B,\cL^{\otimes 3}).
\end{array}\right.
$$
The discriminant and the j-invariant of the corresponding fibration
$Y_C$ are:
$$
\Delta_C\,=\,C^2h^2(h+3k^2)(\gamma^2-h\phi^2)+C^3\ldots,\qquad
j\,\sim\,\frac{1}{C^2}\frac{h^4}{(h+3k^2)(\gamma^2-h\phi^2)}.
$$
The limiting discriminant is
$\Delta_h=h^2(h+3k^2)(\gamma^2-h\phi^2)$, the lowest order term in
$C$ in $\Delta_C$. It is, like $\Delta$, a global section of
$\cL^{\otimes 12}$.

Let $X$ be the hypersurface in the total space of the line bundle
$\cL$ defined by $\xi^2=h$, with 2:1 map $\rho:X\rightarrow B$, and
let $\Delta_0$ be  the divisor in $B$ given by $\Delta_h=0$:
$$
X:\quad \xi^2\,=\,h\quad(\subset\cL),\qquad\qquad \Delta_0:\quad
\Delta_h=0\quad(\subset B).
$$

The preimage in $X$ of $\Delta_0$ is reducible (we omit some of the
$\rho^*$ for simplicity's sake):
$$
\rho^*\Delta_h\,=\,
\xi^4(\xi+\sqrt{-3}k)(\xi-\sqrt{-3}k)(\gamma+\xi\phi)(\gamma-\xi\phi).
$$
The zero locus of $\xi$ in $X$ is the ramification locus of the 2:1
map $\rho:X\rightarrow B$, which is denoted by $O$. The orientifold
in the theory is supported on $O$ and there are two
brane-image-brane pairs.

We denote by $D_1,\ldots,D_r$ the components, with multiplicity one
but not necessarily distinct, of this preimage in $X$  and we denote
their classes by $a_i\rho^*L$. Note that $\sum a_i=12$ since
$\rho^*\Delta_h$ has class $12\rho^*L$. We can take
$D_1=\ldots=D_4=O$, with $a_i=1$, $D_5,D_6$ defined by
$\xi\pm\sqrt{-3}k=0$, also with $a_i=1$, and $D_7,D_8$ defined by
$\gamma\pm\xi\phi=0$ with $a_i=3$. Hence
$$
\rho^*\Delta_0\,=\,D_1+\ldots+D_8,\qquad [D_i]\,=\,a_i\rho^*L,\qquad
(a_1,\ldots,a_8)=(1,1,1,1,1,1,3,3).
$$
In this case, for generic choice of the sections involved, the $D_i$
are smooth. Then one expects the ordinary tadpole relation for the
Euler characteristics of these varieties to hold:
$$
2\chi(Y_C)\,=\,\sum_{i=1}^8\,\chi(D_i),
$$
for any value of $C\in\CC$ for which $Y_C$ is smooth.

Actually in \cite{AE} $\S$4.3.4 a much stronger relation between
Chern classes has been shown to hold. Moreover, this stronger
relation is valid for any choice of the base $B$. Also the $E_7$ and
$E_8$ families as well as the case of singular $D_i$ were considered
in \cite{AE}. In the next section we will determine all the values
of $c_1,d_1,d_2$ for which these stronger, base independent, tadpole
relations from \cite{AE} hold.

\subsection{Numerical tadpole cancellation}\label{tpcc1}
Following \cite{AE} $\S$4.4, we now determine all the cases in which
we have a (numerical) generalized strong tadpole relation:
$$
2\phi_*c(Y)\,=\,\sum_{i=1}^r\,\rho_*c(D_i)
$$
where $Y\subset Z:=\PP E$, $E=\cO_B\oplus \cL^{\otimes d_1}\oplus
\cL^{\otimes d_2}$ is a subvariety with class $[Y]=-K_Z$, so we have
a Calabi-Yau elliptic fibration $\phi:Y\rightarrow B$ and
$\rho:X\rightarrow B$ is a double cover defined by a global section
$h$ of $\cL^{\otimes 2}$ and we assume that $c_1(B)=c_1L$, where $L$
is the class of $\cL$ and
 $c_1$ is a rational number.
We assume that the $D_i$ are nonsingular, with class $a_i\rho^*L$ in
$X$ and that $r\leq 12$, $\sum a_i=12$ and $a_i>0$.

The `generalized' refers to the fact that we do not assume that
$\dim B=3$ nor that the $D_i$ are irreducible components of
$\rho^*\Delta_h$ in a Sen limit. We will see that it may be
impossible to actually find a Sen family which has the $a_i$
required for tadpole cancellation.

The left hand side was computed in equation \ref{PQ} in section
\ref{cec}, the right hand side in section \ref{pf}:
$$
2\frac{P}{Q}c(B)\,=\,\left(\sum_{i=1}^r\,\frac{2a_iL}{1+a_iL}\right)\frac{1+L}{1+2L}c(B).
$$
Here $P,Q$ are polynomials in $L,c_1,d_1,d_2$ multiplied by $c(B)$.
Omitting the factor $c(B)$ on both sides and fixing a partition
$12=a_1+\ldots+a_r$ we obtain, after clearing denominators, a
polynomial in  $L$ with coefficients which are polynomials in
$c_1,d_1,d_2$. Using a computer, we found that these polynomials
have a common zero with $c_1,d_1,d_2\in\QQ$ only if $c_1=1$ and then
there are only the cases {\renewcommand{\arraystretch}{1.3}
$$
\begin{array}{|c|c|c|c|}\hline
(a_1,a_2,\ldots,a_r)&(1,1,1,1,4,4)&(1,1,1,1,1,1,3,3)&(2,2,2,2,2,2)
\\ \hline
(d_1,d_2)&(2,1)&(1,1)&(1,0)\\ \hline
\end{array}
$$
}
(One should note that
$$
\PP(\cO\oplus \cL^{\otimes d_1}\oplus \cL^{\otimes d_2})\,\cong\,
\PP(\cL^{\otimes -d_1}\oplus \cO\oplus \cL^{\otimes
(d_2-d_1)})\,\cong\, \PP(\cL^{\otimes -d_2}\oplus \cL^{\otimes
(d_1-d_2)}\oplus \cO)
$$
and we list only one of these cases.) In the three cases we listed,
the generalized strong tadpole relation is universally satisfied,
that is, we omitted the factors $c(B)$ on both sides. Such relations
thus imply generalized strong tadpole relations for any base $B$.

The first two cases were already found in \cite{AE} $\S$ 4.3.3 and
4.3.4, they occur in the $E_7$ and $E_6$ case respectively. In
particular, the other case listed in their Theorem 4.9 does not
occur for the elliptic fibrations we consider here (that is,
$2\phi_*c(Y)\neq 12L/(1+2L)c(B)$ for the $Y$ we considered). This
case thus remains without a geometric interpretation.

It is in fact trivial to verify the universal strong tadpole
cancellation in the new $(1,0)$-case: if $c_1=d_1=1$, $d_2=0$ then,
up to a common factor, $P=12L(1+L)$, $Q=(1+2L)^2$ and
$$
2\phi_*c(Y)\,=\, \frac{24L(1+L)}{(1+2L)^2}c(B)\,=\,
\frac{24L}{1+2L}\cdot \frac{1+L}{1+2L}c(B)\,=\, 6\cdot\frac{2\cdot
2L}{1+2L}\cdot \frac{1+L}{1+2L}c(B).
$$

\subsection{No geometrical example?}
In section \ref{e6tp} we copied the Sen limit with
$(c_1,d_1,d_2)=(1,1,1)$, the $E_6$-case, which realizes
$(1,\ldots,1,3,3)$, from \cite{AE}. Now we want to find a Sen limit
in the $(c_1,d_1,d_2)=(1,0,1)$ case which realizes
$(a_1,\ldots,a_6)=(2,\ldots,2)$. The main issue is that in the
examples we tried, the limiting discriminant $\Delta_h$ always has a
factor $h$:
$$
\Delta_h\,=\,h^a\underline{D},\qquad\mbox{so}\quad
\rho^*\Delta_h\,=\,\xi^{2a}\rho^*\underline{D},
$$
with $a>0$ and a section $\underline{D}$ of $\cL^{\otimes(12-2a)}$.
This implies that the components of the preimage of $\Delta_h=0$ in
$X$ has components $D_1=\ldots=D_{2a}=O$ with multiplicity one. As
$[D_i]=[O]=1\cdot \rho^*L$ this gives $a_1,\ldots,a_{2a}=1$. Thus it
seems impossible to get a limit with all $a_i=2$.

In a Weierstrass model $y^2=x^3+fx+g$, this can be seen rather
easily. A Sen family would be given by $f=f_0+f_1C+f_2C^2+\ldots$,
$g=g_0+g_1C+g_2C^2+\ldots$. Since for $C=0$ one requires
$\Delta=4f^3+27g^2=0$, one can put (at least locally) $f_0=-3h^2$,
$g_0=-2h^3$. But then $\Delta =h^2\Delta_1(h,f_1,g_1)C+\ldots$, so
if $\Delta_1$ is not identically zero on the base, we get
$\Delta_h=h^2\Delta_1(h,f_1,g_1)$ and thus $a\geq 2$. If $\Delta_1$
is identically zero, one finds conditions on $f_1,g_1$ which then
imply that  $\Delta=h^a\Delta_2(h,\ldots)C^2+\ldots$ with $a>0$ etc.
See also \cite{Ketc} for transformations to the Weierstrass model
and the (local) divisibility argument.

\

\section*{acknowledgments}

We acknowledge T. Weigand, T. W. Grimm and J. Fullwood for useful comments.\\
This work was partially supported by INFN.

\normalsize

\

\appendix
\section{Sen's weak coupling limit revisited}
In this appendix we collect some observations on Sen limits. The
weak coupling limit, also called a Sen limit, of an elliptic
fibration $Y$ over a base $B$ is a family of elliptic fibrations
$Y_C\rightarrow B$, with $C\in\CC$, such that $Y$ is a deformation
of the general $Y_C$ and such that the general fiber of $Y_0$ has
$j$-invariant $\infty$.

\subsection{The limit fibration $Y_0$}
The simplest limit fibrations $Y_0$ already present a non-trivial
geometry. In particular, an auxiliary Calabi-Yau manifold on which
the type IIB theory is usually studied, appears naturally.

The Weierstrass model of an elliptic fibration, its discriminant and
j-invariant are:
$$
y^2\,=\,x^3\,+\,fx\,+\,g,\qquad
\Delta\,=\,4f^3\,+\,27g^2\,=\,0,\qquad j=4(24f)^3/\Delta.
$$
Here $f,g$ are global sections of $\cL^{\otimes 4},\cL^{\otimes 6}$
for some line bundle $\cL$ on the base. In case $\cL=\omega_B^{-1}$,
the anti-canonical bundle, one obtains Calabi-Yau varieties.

A limit fibration has $j(b)=\infty$ and thus has $f(b)\neq 0$,
$\Delta(b)=0$ for the general $b\in B$. The simplest way to get such
limits is to take a section $h$ of $\cL^{\otimes 2}$ and to put
$$
f\,:=\,-3h^2,\quad g\,:=\,-2h^3.
$$
We will consider this limit fibration $Y_0$ in the remainder of this
section,
$$
Y_0:\qquad y^2\,=\,x^3\,-3h^2x\,-2h^3.
$$
Notice that $x^3-3h^2x-2h^3=(x+h)^2(x-2h)$.

The first observation we'd like to make is that these limits are
quotients of a product $X\times E_\infty$, here $X$ is double cover
of $B$ and $E_\infty$ is a singular cubic curve with a node in the
point $(x,y)=(-1,0)$:
$$
E_\infty:\quad y^2=x^3-3x-2,\qquad(\mbox{note}\quad
x^3-3x-2=(x+1)^2(x-2)).
$$

In fact, the fiber of $Y_0$ over any point $b\in B$ with $h(b)\neq
0$ is isomorphic to $E_\infty$. However, $Y_0$ is not the product of
$B\times E_\infty$ (and is not even birational to this product if
$h$ is not a square), due to a global twist. In fact, for $b\in B$
with $h(b)\neq 0$, the fiber is nodal cubic cubic with equation
$y^2=-3h(b)u^2+u^3$, where $u=x+h(b)$. Thus the two tangent lines at
the node are $y=\pm \sqrt{-3h(b)}u$ and there is a non-trivial
monodromy around $h=0$ if $h$ is not a square. Note that $h=0$ is
also the locus where the fibers are no longer nodal, but are
cuspidal, so they are not stable anymore. This aspect is emphasized
in \cite{AE}, who single out the points where $\Delta(b)=f(b)=0$.

We trivialize the monodromy on the double cover  of the base $B$,
branched over $h=0$, defined by
$$
X:\quad \xi^2\,=\,h,\qquad \rho:X\longrightarrow B,\quad
(\xi,b)\,\longmapsto\,b.
$$
This double cover naturally lives inside the total space of the line
bundle $\cL$. If $h=0$ is a smooth subvariety of $B$, then $X$ is
also smooth and $X$ is a Calabi-Yau variety if $\cL=\omega_B^{-1}$.

Now we pull-back the fibration $Y_0$ along $\rho$ to $X$, so we
consider the fibration defined by the Weierstrass equation
$y^2=x^3-3h^2x-2h^3$ as a fibration $\tilde{Y}_0$ over $X$. This
fibration is independent of $\xi$, but on $X$ we can substitute
$h:=\xi^2$. Doing so, and substituting also
$$
x\,:=\xi^2u,\quad y:=\xi^3v,\quad h\,:=\,\xi^2,\quad\mbox{we
get}\qquad v^2\,=\,u^3\,-\,3u\,-\,2.
$$
So we see that, at least over the open subset of $X$ where $\xi\neq
0$, the pull-back $\tilde{Y}_0$ is isomorphic to the product
$X\times E_\infty$.
$$
\xymatrix{ {X \times E_\infty} \ar @{-->}[r]^{\phantom{XX}\approx}
\ar[dr] & \tilde{Y}_0 \ar[d]  \ar[r]& Y_0 \ar[d]\\
 & X \ar[r]^{} & B
}
$$
Conversely, starting from the double cover $X$ and the nodal cubic
$E_\infty$, we can recover $Y_0$ as a quotient of $X\times E_\infty$
by the action of an involution $\tilde{\sigma}$. The covering group
of $\rho:X\rightarrow B$ has only one non-trivial element
$$
\sigma:\,X\,\longrightarrow\,X,\qquad
(\xi,b)\,\longmapsto\,(-\xi,b).
$$
Now we define an automorphism, of order two,
$$
\tilde{\sigma}\,:X\times E_\infty\,\longrightarrow\,X\times
E_\infty,\qquad ((\xi,b),(u,v))\,\longmapsto\,((-\xi,b),(u,-v)).
$$
The non-trivial invariants for this action are $\xi^2=h$,
$v^2=u^3-3u-2$ and $t:=\xi v$. Thus the quotient is the fibration on
$B$ defined by $(\xi v)^2=h(u^3-3u-2)$. It can be put in Weierstrass
form by multiplying both sides by $h^2$ and defining $y:=h\xi
v=\xi^3v$ and $x:=hu$. This gives the equation of $Y_0$, so we
established a birational isomorphism (not well-defined where $h=0$
for example):
$$
(X\times E_\infty)/\tilde{\sigma}\,\simeq\,Y_0.
$$
The non-semistable fibers, i.e.\ those with $\Delta=f=0$, arise from
the fixed points in the action of $\tilde{\sigma}$. These are the
cuspidal fibers in the Weierstrass model and they occur over the
$b\in B$ with $h(b)=0$.

A second observation is that the points $((\xi,b),(2,0))\in X\times
E_\infty$ map to a section of the quotient fibration, given by
$$
s_2:\,B\,\longrightarrow\,Y_0,\qquad b\,\longmapsto\,(b,(x=2h,y=0)),
$$
in fact $y^2=x^3-3h^2x-2h^3=(x+h)^2(x-2h)$. This section has order
two in the Mordell-Weil group of $Y_0$ over $B$ (recall that the
smooth points of a nodal cubic form a group isomorphic to the
multiplicative group of complex numbers, \cite{Cas}, $\S$9.).

\subsection{Sen's weak coupling limit}
Starting from $Y_0$ and an elliptic fibration $Y$ over $B$ in
Weierstrass form $y^2=x^3+fx+g$ one can obtain a weak coupling limit
simply by defining $Y_C$ to be the fibration defined by
$y^2=x^3+(Cf-3h^2)x+(Cg-2h^3)$. Unfortunately, the physics of this
general limit is hard to understand.

In Sen's limit, one considers the simpler family:
$$
f\,:=\,C\eta,\qquad \,g:=\,Ch\eta\qquad\mbox{so}\quad Y_C:\quad
y^2\,=\,x^3+(C\eta-3h^2)x+(Ch\eta-2h^3).
$$
A consequence of this choice is that the cubic polynomial factors:
$$
x^3+(C\eta-3h^2)x+(Ch\eta-2h^3)\,=\,(x+h)(x^2 -hx + C\eta - 2h^2).
$$
Thus Sen's family has a section, of order two in the Mordell-Weil
group, given by $b\mapsto (x,y)=(-h,0)$. This section specializes to
the node in the fibers over $C=0$.

\subsection{Global aspects of Sen's limit}
We now consider some global aspects of Sen's family, which we view
as an elliptic fibration defined by the Weierstrass model above. To
simplify the computations, we change the coordinates by putting
$x:=x+h$, so the order two section becomes $x=0$ and Sen's family
$Y$ is defined by the Weierstrass equation
$$
Y:\qquad y^2\,=\,x(x^2-3hx+C\eta).
$$
We consider $Y$ as an elliptic fibration over $B\times\CC$. It has a
natural compactification in the bundle of projective planes $\PP(
\cL^{\otimes 2}\oplus \cL^{\otimes 3}\oplus \cO_B)$ over
$B\times\CC$, by homogenizing the equation to a cubic in $x,y,z$.

The total space $Y$ of this fibration is singular in general. For
example, the points where $x=y=\eta=0$ are easily seen to be
singular on $Y$. Over $C=0$ we have additional singular points which
are due to the fact that for each $b\in B$ the section $(x,y)=(0,0)$
meets the node in the fiber over $b$. Since the limit $C\rightarrow
0$ is the most important feature of this family, we show how to get
a better model.

Put $x:=Cx$, $y:=Cy$ in the equation for $Y$ and divide the result
by $C^2$. This gives, after homogenization, the equation
$$
Y':\qquad y^2z=x(Cx^2-3hxz+\eta z^2)
$$
which defines an elliptic fibration $Y'$ over $B\times\CC$.
Obviously, $Y$ and $Y'$ are isomorphic over $B\times
(\CC-\{0\})$,where $C\neq 0$. However, the fibers over points
$(b,C)$ with $C=0$ are now reducible cubic curves, consisting of a
conic $y^2=-3h(b)x^2+\eta(b) xz$ and the line at infinity $z=0$.
These are fibers of type $I_2$ if $(h(b),\eta(b))\neq (0,0)$. The
points of $Y'$ over $(b,C)$ with $b$ general in $B$ and $C=0$ are
smooth on $Y'$, so this is a better model than $Y$.

\subsection{A toy model}
A key feature of Sen's limit is the presence of $-T^{-4}$-monodromy
over a path which encloses two $I_1$-fibers. This monodromy
indicates the presence of a $D7$ brane in  the limit. Here we
consider a Sen limit of a family of elliptic fibrations
$Y_C\rightarrow B=\PP^1$, for $C\in\CC$, where this monodromy is
easily visible. In fact, each $Y_C$, for $C\neq 0$, has only three
bad fibers. Over $s=\pm 2/3\sqrt{C}$ where $s$ is the parameter on
the base $\PP^1$, we have $I_1$ fibers, that is nodal cubic curves,
whereas over $s=\infty$ we have an $I_4^*$-fiber. Thus the local
monodromy around infinity is $-T^4$, but a path around infinity is
also a path, transversed in the opposite direction, which encloses
the two points $s=\pm  2/3\sqrt{C}$. Thus the monodromy over a path,
transversed in the positive direction, which encloses these two
points will be $(-T^4)^{-1}=-T^{-4}$.

The example is as follows. Let $(s:t)$ be the homogeneous
coordinates on $\PP^1$ and consider the rational elliptic fibration
given by the Weierstrass model
$$
Y_C:\quad y^2\,=\,x^3+(Ct^4-3(st)^2)x+(Cst^5-2(st)^3),\qquad
\Delta\,=\,C^2t^{10}(4Ct^2-9s^2)
$$
Here $\Delta=4f^3\,+\,27g^2$ is the discriminant of
 $y^2=x^3+fx+g$. The j-invariant is
$$
j=\frac{4(24f)^3}{\Delta}=\frac{4(24(Ct^4-3(st)^2))^3}{C^2t^{10}(4Ct^2-9s^2)}.
$$
From the order of vanishing of $\Delta$ and $j$ one deduces
immediately that for $C\neq 0$ the fibers over $s=\pm 2/3\sqrt{C}t$
are of type $I_1$ and that the fiber over $t=0$ (i.e.\ $s=\infty$)
is of type $I_4^*$.

A peculiar feature of this example is that the $Y_C$, for $C\neq 0$,
are all isomorphic. In fact, we have isomorphisms
$$
\phi_C:\,Y_1\,\stackrel{\cong}{\longrightarrow}\,Y_C,\qquad
((x,y),(s:t))\,\longmapsto
\Bigr((\sqrt{C}x,\sqrt{C^3}y),(s:\sqrt{C}t)\Bigl).
$$
The existence of such a family of fibrations is related to the finer
points of the moduli theory of Weierstrass models and is discussed
in \cite{Mw}.

\

\


\begin{thebibliography}{PQR}
\bibitem[AE1]{AE2} P.\ Aluffi, M.\ Esole,
{\it Chern class identities from tadpole matching in type IIB and
F-theory}, (arXiv:0710.2544), JHEP {\bf 03} (2009) 032.

\bibitem[AE2]{AE} P.\ Aluffi, M.\ Esole,
{\it New Orientifold Weak Coupling Limits in F-theory},
(arXiv:0908.1572v3), JHEP 1002:020, 2010.

\bibitem[BHV1]{BHV}
C.\ Beasley, J.\ J.\ Heckman and C.\ Vafa, {\it GUTs and Exceptional
Branes in F-theory - I}, JHEP {\bf 0901} (2009) 058.

\bibitem[BHV2]{BHV1}
C.\ Beasley, J.\ J.\ Heckman and C.\ Vafa, {\it GUTs and Exceptional
Branes in F-theory - II: Experimental Predictions}, JHEP {\bf 0901}
(2009) 059.

\bibitem[BHT]{BHT}
A.\ P.\ Braun, A.\ Hebecker, H.\ Triendl, {\it D7-Brane Motion from
M-Theory Cycles and Obstructions in the Weak Coupling Limit}, Nucl.\
Phys.\  {\bf B800 } (2008)  298-329.

\bibitem[BGJW]{GW}
  R.\ Blumenhagen, T.\ W.~Grimm, B.\ Jurke, T.\ Weigand,
  {\it Global F-theory GUTs},
  Nucl.\ Phys.\  {\bf B829 } (2010)  325-369.

\bibitem[C]{Cas} J.W.S.\ Cassels, { Lectures on elliptic curves}.
London Mathematical Society Student Texts 24. Cambridge University
Press 1991.

\bibitem[D]{D}
F.\ Denef, {\it Les Houches Lectures on Constructing String Vacua},
arXiv:0803.1194 [hep-th].

\bibitem[DMSS]{DMSS}
M.J.\ Dolan, J.\ Marsano, N.\ Saulina, S.\ Schafer-Nameki, {\it
F-theory GUTs with U(1) Symmetries: Generalities and Survey},
 arXiv:1102.0290.

\bibitem[EY]{EY} M.\ Esole, S-T.\ Yau,
{\it Small resolutions of SU(5)-models in F-theory},
arXiv:1107.0733.

\bibitem[FJ]{FJ}
J.\ Fullwood, {\it On generalized Sethi-Vafa-Witten formulas},
[arXiv:1103.6066 [math.AG]]; to appear on Journal of Mathematical
Physics.


\bibitem[F]{F} W.\ Fulton, { Intersection Theory}. Springer-Verlag, Berlin 1984.

\bibitem[GGO]{GGO} A.\ Grassi, Z.\ Guralnik, B.\ Ovrut,
{\it Five-brane BPS states in heterotic M-theory}, JHEP {\bf 0101 }
(2001)  037.

\bibitem[GKW]{GW1}
  T.~W.~Grimm, S.~Krause, T.~Weigand,
  {\it F-Theory GUT Vacua on Compact Calabi-Yau Fourfolds},
  JHEP {\bf 1007 } (2010)  037.

\bibitem[GW]{GW2}
  T.\ W.\ Grimm, T.\ Weigand,
  {\it On Abelian Gauge Symmetries and Proton Decay in Global F-theory GUTs},
  Phys.\ Rev.\  {\bf D82 } (2010)  086009.

\bibitem[H]{H}
J.\ J.\ Heckman, {\it Particle Physics Implications of F-theory},
arXiv:1001.0577 [hep-th].

\bibitem[HKSV]{HKSV}
J.\ J.\ Heckman, G.\ L.\ Kane, J.\ Shao and C.\ Vafa, {\it The
Footprint of F-theory at the LHC}, JHEP {\bf 0910} (2009) 039.

\bibitem[HTV]{HTV}
J.\ J.\ Heckman, A.\ Tavanfar and C.\ Vafa, {\it Cosmology of
F-theory GUTs}, JHEP {\bf 1004} (2010) 054.

\bibitem[HV]{HV}
J.\ J.\ Heckman and C.\ Vafa, {\it From F-theory GUTs to the LHC},
arXiv:0809.3452 [hep-ph].

\bibitem[KMSS]{Ketc}S.\ Katz, D.\ R. Morrison, S. Sch\"afer-Nameki, J.\ Sully,
{\it Tate's algorithm and F-theory}, arXiv:1106.3854.

\bibitem[MSS1]{MSS1} J.\ Marsano, N.\ Saulina, S.\ Sch\"afer-Nameki,
{\it Compact F-theory GUTs with $\rm U(1)_{PQ}$}. JHEP 1004:095,2010
(arXiv:0912.0272).

\bibitem[MSS2]{MSS2} J.\ Marsano, N.\ Saulina, S.\ Sch\"afer-Nameki,
{A Note on G-Fluxes for F-theory Model Building}, JHEP 1011:088,2010
(arXiv:1006.0483).

\bibitem[M1]{Mw} R.\ Miranda,
{\it The Moduli of Weierstrass Fibrations Over P1}, Math.\ Ann.\
{\bf 255} (1981) 379--394.

\bibitem[M2]{M} R.\ Miranda, {\it Smooth models for elliptic threefolds},
in: The birational geometry of degenerations (Cambridge, Mass.,
1981), pp. 85--133, Progr.\ Math., 29, Birkh\"auser, Boston, Mass.,
1983.

\bibitem[S1]{S1}
A.\ Sen, {\it F theory and orientifolds}, Nucl.\ Phys.\  B {\bf 475}
(1996) 562.

\bibitem[S2]{S} A.\ Sen,
{\it Orientifold Limit of F-Theory Vacua}, Phys.\ Rev.\ D (3)  {\bf
55}  (1997) 7345--7349 (arXiv:hep-th/9702165).

\bibitem[V]{V}
C.\ Vafa, {\it Evidence for F theory}, Nucl.\ Phys.\  B {\bf 469}
(1996) 403.


\end{thebibliography}
\end{document}